\documentclass[aps,prl,preprint,superscriptaddress]{revtex4}

\usepackage{graphicx}

\usepackage{amssymb}

\begin{document}

\title{Effect of light Sr doping on the spin--state transition in
LaCoO$_3$}

\author{A.~Podlesnyak}
\affiliation{Hahn--Meitner--Institut, SF--2, Glienicker Stra{\ss}e 100, Berlin 141 09, Germany}
\affiliation{Laboratory for Neutron Scattering, ETH
Z\"urich $\&$ Paul Scherrer Institut, CH-5232 Villigen PSI,
Switzerland}
\author{K.~Conder}
\affiliation{Laboratory for Developments and Methods, Paul Scherrer Institute,
CH-5232 Villigen PSI, Switzerland}
\author{E.~Pomjakushina}
\affiliation{Laboratory for Developments and Methods, Paul Scherrer Institute,
CH-5232 Villigen PSI, Switzerland}
\affiliation{Laboratory for Neutron Scattering, ETH
Z\"urich $\&$ Paul Scherrer Institut, CH-5232 Villigen PSI,
Switzerland}
\author{A.~Mirmelstein}
\affiliation{Institute for Metal Physics RAS, GSP--170, 620041 Ekaterinburg, Russia}
\author{P.~Allenspach}
\affiliation{Laboratory for Developments and Methods, Paul Scherrer Institute,
CH-5232 Villigen PSI, Switzerland}
\author{D.~I.~Khomskii}
\affiliation{II~Physikalisches Institut, Universit{\"a}t zu K{\"o}ln,
Z{\"u}lpicher Stra{\ss}e 77, 50937 K{\"o}ln, Germany}

\begin{abstract}
We present an inelastic neutron scattering study of the low energy
crystal--field excitations in the lightly doped cobalt perovskite
La$_{0.998}$Sr$_{0.002}$CoO$_3$. In contrast to the parent compound
LaCoO$_3$ an inelastic peak at energy transfer $\sim 0.75$~meV was found at
temperatures below 30~K. This excitation apparently corresponds to a
transition between a ground state orbital singlet and a higher excited
orbital doublet, originating from a \emph{high-spin triplet} split by a
small trigonal crystal field. Another inelastic peak at an energy transfer
$\sim 0.6$~meV was found at intermediate temperatures starting from $T >
30$~K. This confirms the presence of a thermally induced spin--state
transition from the low--spin Co$^{3+}$ to a magnetic high--spin state in
the non-disturbed LaCoO$_3$ matrix. We suggest that hole doping of
LaCoO$_3$ leads to the creation of a magnetic polaron and hence to the
low--to--high spin state transition on the relevant Co sites.
\end{abstract}

\maketitle

The magnetic and transport properties of LaCoO$_3$ have peculiar
temperature dependencies due to thermally induced changes in the
electronic configuration of Co$^{3+}$ ions \cite{Raccah,Rao}. The
ground state is low spin (LS), t$_{2g}^6$, S=0. With increasing
temperature, first a crossover into a magnetic, but still insulating
state, appears at about 80-120~K, followed by another crossover into
a "bad metallic", magnetic state at $T\sim400-600$~K. Such an
anomalous evolution of the spin states with temperature is debated
already for the last 50 years.

Hole-doped lanthanum cobalt oxides La$_{1-x}$Sr$_x$CoO$_3$ show a
spin-glass ($x<0.18$), a cluster-glass ($0.18<x<0.5$) and a
ferromagnetic metallic state ($x>0.5$) at low temperatures
\cite{Itoh,Good}. So far, most of the investigations are
concentrated on the middle- and high-level doped system
La$_{1-x}$Sr$_x$CoO$_3$, $x>0.1$. However, already the lightly doped
material $x\sim0.002$ (i.e with the estimated concentration of two
holes per thousand Co$^{3+}$ ions) exhibits paramagnetic properties
at low temperatures, in strong contrast to the parent diamagnetic
insulator LaCoO$_3$ \cite{Yamaguchi}. Recently, using the inelastic
neutron scattering (INS) technique, we identified the energy levels
of the thermally excited states of Co$^{3+}$ ions in LaCoO$_3$
\cite{Podlesnyak}. In the present work we performed INS measurements
in the lightly doped cobalt perovskite
La$_{0.998}$Sr$_{0.002}$CoO$_3$ in order to obtain information on
the ground and low-lying excited states of Co ions.

\begin{figure}[t]
\begin{center}
\includegraphics [width=0.8\columnwidth]{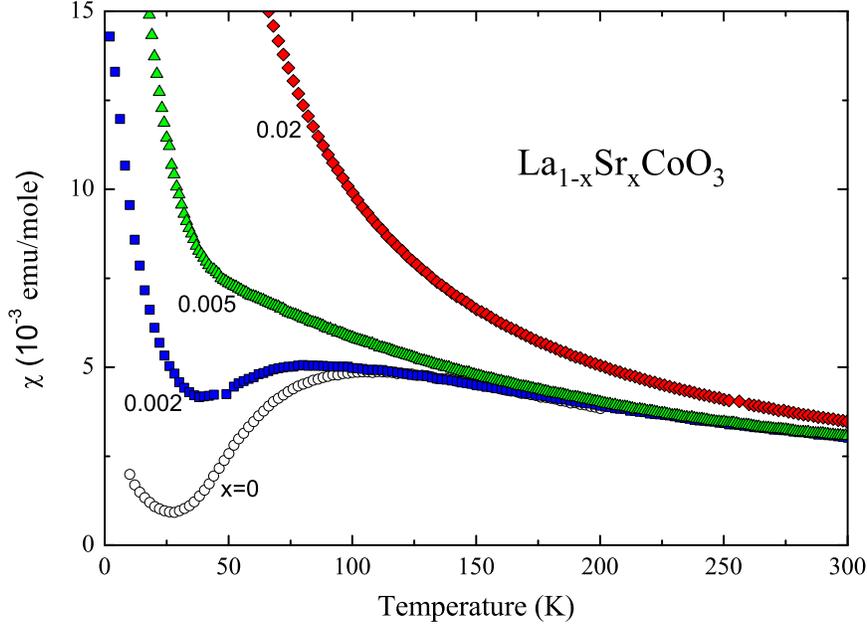}
\caption{Temperature dependence of the magnetic
susceptibility of lightly doped La$_{1-x}$Sr$_x$CoO$_3$ taken at 1~T.
\label{fig1}}
\end{center}
\end{figure}

The samples of La$_{1-x}$Sr$_x$CoO$_3$ ($x=$~0, 0.002, 0.005) were
synthesized by a solid state reaction using La$_2$O$_3$, SrCO$_3$
and Co$_3$O$_4$ of a minimum purity of 99.99\%. Each polycrystalline
sample was prepared in amount of 50~g, therefore the absolute mass
of the doping compound was sufficient to weight it with high
accuracy. The phase purity of the synthesized compounds were checked
with an x-ray diffractometer (SIEMENS D500). The crystal growth
experiments were carried out using an Optical Floating Zone Furnace
(FZ-T-10000-H-IV-VP-PC, Crystal System Corp., Japan). Oxygen content
of polycrystalline and single crystal samples was determined by
thermogravimetric hydrogen reduction \cite{Conder}. For all the
samples oxygen stoichiometry was found to be $3.00\pm0.01$. The
small concentration of the doping element made it difficult to
control the distribution of strontium along the grown crystals with
traditional techniques such as EDX. Instead, we have compared
temperature dependencies of the magnetization of crystal pieces
taken from different places of the crystal (see Fig.~\ref{fig1}).
The results of the magnetization measurements obtained for all the
crystal pieces and also for starting powder were identical (within
each strontium concentration) and consistent with previously
published data \cite{Yamaguchi}. The INS measurements were performed
for both polycrystalline and single crystal samples on the
high-resolution time-of-flight spectrometer FOCUS \cite{focus}
 in the temperature interval of $1.5-100$~K.
The data were collected using an incoming neutron energy of 3.5~meV,
resulting in an energy resolution at the elastic position [full
width at half maximum (FWHM)] of 0.1~meV.

\begin{figure}[tb!]
\begin{center}
\includegraphics [width=0.65\columnwidth]{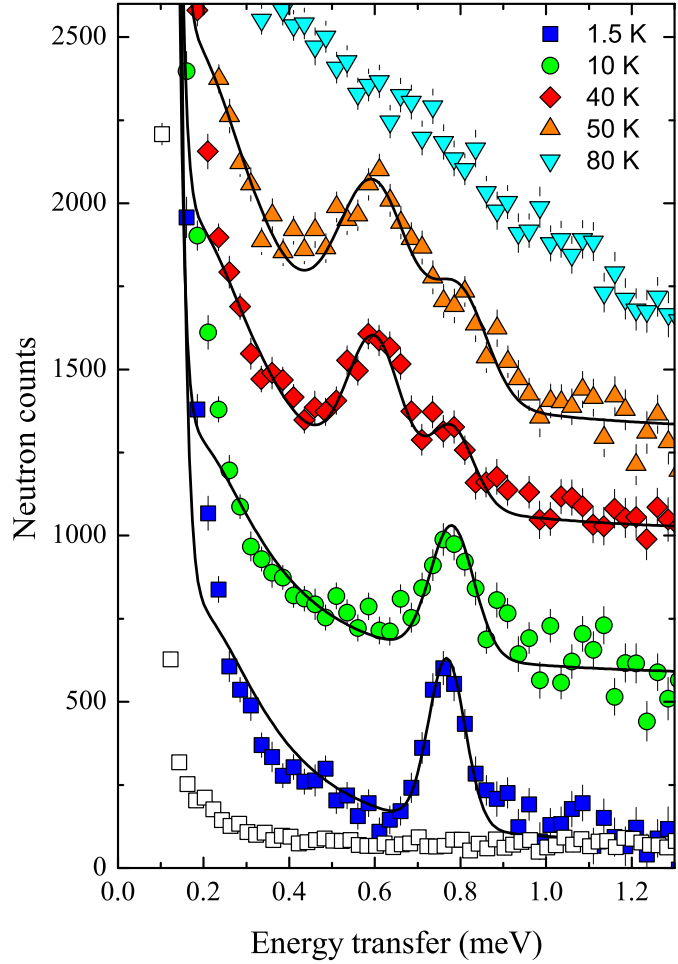}
\caption{Temperature evolution of the INS profiles
measured in La$_{0.998}$Sr$_{0.002}$CoO$_3$. The open squares
correspond to the LaAlO$_{3}$ nonmagnetic reference compound. The
lines are guides to the eye. For clarity, an offset has been added
to the various curves.
\label{fig2}}
\end{center}
\end{figure}

In contrast to the parent compound LaCoO$_{3}$, where no excitations
have been found for temperatures $T<30$~K \cite{Podlesnyak}, an
inelastic peak at energy transfer $\delta E_1 \sim 0.75$~meV was
observed already at $T=1.5$~K (see Fig.~\ref{fig2}). This feature is
apparently intrinsic and not related to surface properties: we
obtained practically the same results on a single crystal, although
with poorer statistics due to the small size (6~g) of our
single--crystalline sample. One more inelastic peak $\delta E_2 \sim
0.6$~meV was found at intermediate temperatures starting from $T
\sim 30$~K. A strong broadening of the transitions was observed with
increasing temperature.

The $\delta E_2$ excitation is due to a thermally excited magnetic
state of Co$^{3+}$ ions in the non-disturbed LaCoO$_3$ matrix, as
was discussed in our previous work \cite{Podlesnyak}. The first
magnetic excited state in LaCoO$_{3}$ can be a high spin (HS) state
with a small non-cubic crystal field, or it can be an intermediate
spin (IS) state with a large non-cubic crystal field or orbital
ordering. There is one striking difference between these scenarios:
this is the predicted $g$-factor. The HS state with a small
non-cubic distortion is a spin--orbit triplet with a $g$-factor of
about 3.5, whereas the IS with strong non-cubic distortion is a spin
triplet with a $g$-factor of about 2.0. Since the $g$-factor
obtained from the field dependence of the INS is $g\sim3$
\cite{Podlesnyak}, the HS scenario seems to us more plausible.
Following the proposed model \cite{Podlesnyak}, we suggest that the
$\delta E_1$ peak also corresponds to the transition between an
orbital singlet and an orbital doublet originating from the HS
triplet, split by a trigonal crystal field. The doped hole is the
cause for the LS$\rightarrow$HS state transition on the relevant Co
site. However, the integral intensities of the $\delta E_1$ and
$\delta E_2$ excitations are comparable. This means that the
intensity of the $\delta E_1$ excitation is much higher than what
would follow from an estimated concentration of doped holes per Co
site ($x=0.002$). Therefore, we propose that holes introduced in the
LS state of LaCoO$_3$  by replacing the trivalent La$^{3+}$ ions
with divalent Sr$^{2+}$ ions are extended over the neighboring Co
sites, forming thus magnetic polaron and transforming all the
involved Co ions (e.g. 8 per each Sr) to the high-spin state.

The authors thank D. N. Argyriou for fruitful
discussions. This work is based on experiments performed at the
Swiss spallation neutron source SINQ, Paul Scherrer Institute,
Villigen, Switzerland. We are indebted to the Swiss National Science
Foundation for financial support through grant SCOPES
IB7320-110859/1 and NCCR MaNEP project.

\end{document}